\documentclass[unsortedaddress,twocolumn,pre,amsmath,amssymb]{revtex4}

\usepackage{graphicx}
\usepackage{dcolumn}
\usepackage{bm}
\begin{document}

\preprint{APS/123-QED}

\title{Achievement of Alternative Configurations of Vehicles on Multiple Lanes}
\author{Ryosuke Nishi$^{*}$, Hiroshi Miki$^{\dagger }$, Akiyasu Tomoeda$^{*}$, and Katsuhiro Nishinari$^{* \ddagger }$}
\affiliation{
$^{*}$ Department of Aeronautics and Astronautics,
Faculty of Engineering, University of Tokyo,
Hongo, Bunkyo-ku, Tokyo 113-8656, Japan.
\\
$^{\dagger}$ SAKURA ACADEMIA Co., JAPAN.
\\
$^{\ddagger}$ PRESTO, Japan Science and Technology Corporation,
Hongo 7-3-1, Bunkyo-ku, Tokyo, 113-8656, JAPAN.
}

\date{\today}
\begin{abstract}
Heavy traffic congestion daily occurs at merging sections on highway. 
For releasing this congestion, possibility of alternative configuration of vehicles on multiple-lane road is discussed in this paper.
This is the configuration where no vehicles move aside on the other lane.
It has a merit in making smooth merging at an intersection or a junction due to so-called the "zipper effect". 
We show, by developing a cellular automaton model for multiple lanes, 
that this configuration is achieved by simple local interactions between vehicles neighboring each other. 
The degree of the alternative configuration in terms of the spatial increase of parallel driving length is studied 
by using both numerical simulations and mean field theory. 
We successfully construct a theoretical method for calculating this degree of the alternative configuration by using cluster approximation. 
It is shown that the theoretical results coincide with those of the simulations very well.
\end{abstract}
\maketitle
\section{INTRODUCTION}
In recent years, traffic dynamics has attracted much interest of physicists, 
and thus it has been studied more and more diligently \cite{D.Helbing} \cite{D.Chowdhury}.
Researchers have mainly developed the analysis of the traffic flow on one-lane road 
by using continuous models \cite{M.Bando} and cellular automaton (CA) models \cite{M.R.Evans}. 
Recently, the analysis of the traffic flow on multiple-lane road at an intersection or a junction is considered 
as one of the important study of traffic flow for releasing traffic congestion. 
Kita modeled merging interactions with game theory \cite{H.Kita}.
Hidas investigated vehicle interactions in merging and weaving by using agent based simulations \cite{P.Hidas}.
Davis introduced the cooperation in merging by adding interactions between pairs of vehicles in opposite lanes \cite{Davis}.  
He showed that velocity of vehicles in cooperative merging was higher than that in no cooperation.

However, these previous works were qualitative and did not study the configuration of vehicles in detail 
on two lanes before an intersection or a junction which determines the efficiency of merging. 
Among various configurations, the alternative configuration is the best 
because it realizes the smooth "zipper" merging, which is the merging of vehicles by turns on one lane and on the opposite lane. 
Thus, the transformation of the configuration toward the alternative configuration is significant for the improvement of the congested flow at a junction. 

The purpose of this paper is to propose a simple and natural method for achievement of the alternative configuration of vehicles on two lanes. 
Then we study the method by both computer simulations and a mean field theory. 
The communication of vehicles on two lanes is studied by the CA model which we call the multiple lanes stochastic optimal velocity (MLSOV) model.  
This is an extension of SOV model proposed in \cite{M.Kanai} 
by introducing the interactions between vehicles on the opposite lane.

This paper is organized as follows. 
In Sec. I\hspace{-.1em}I we define MLSOV model, 
and results of simulations of two-lane flow before an intersection is presented in Sec. I\hspace{-.1em}I\hspace{-.1em}I. 
Then in Sec. I\hspace{-.1em}V we calculate the degree of alternative configuration by using cluster approximation 
and compare it with the results of the simulations.
Sec. V is devoted to the concluding discussions.
\section{A Multiple Lanes CA Model Without Lane Change}
\begin{figure}[htbp]
\includegraphics[width=1.0\linewidth]{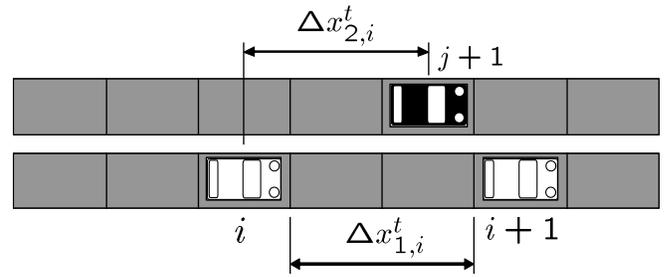}
\caption{
The vehicles that affect the behaviour of $i$-th vehicle in MLSOV model. 
$i+1$-th vehicle is at the cell which is $\Delta x_{1,i}^{t}$ cells ahead of $i$-th vehicle on the same lane. 
$j+1$-th vehicle is the closest vehicle to the $i$-th vehicle with the distance $\Delta x_{2,i}^{t}$ cells ahead on the neighboring lane.
}
\label{fig:mlsov_model}
\end{figure}

We propose here the MLSOV model by introducing interactions between two lanes.
We choose the SOV model as a basis of our extended model because it is one of the simplest model whose fundamental diagram, i.e. flow versus density plot, 
shows the metastable state observed in real traffic data \cite{M.Kanai}. 
MLSOV model represents the interactions of vehicles on two lanes 
by seeing the ones on the opposite lane each other.
Note that we do not take into account of the lane change behaviour in this paper 
because we focus only on the achievement of the alternative configuration of vehicles before an intersection. 
In the model, the movement of all vehicles on both lane $1$ and lane $2$ is ruled as follows. 
$i$-th vehicle on each lane moves one cell in front in one time step with probability $v_{i}^{t}$ at time $t$ 
provided that the next cell is empty. 
Thus the movement is described as 
\begin{align}
x_{i}^{t+1} = \begin{cases}
              x_{i}^{t}+1, & \text{with probability $v_{i}^{t}$} \\
              x_{i}^{t},   & \text{with probability $1-v_{i}^{t}$,}
              \end{cases}
\label{eq:x_i_t+1}
\end{align}
where $x_{i}^{t}$ is the position of $i$-th vehicle at time $t$ as shown in figure \ref{fig:mlsov_model}. 
$v_{i}^{t}$ is called intension and normalized as $0 \le v_{i}^{t} \le 1$. 
$i+1$-th vehicle is located at the cell which is $\Delta x_{1,i}^{t}$ cells ahead of $i$-th vehicle on the same lane 
as shown in figure \ref{fig:mlsov_model}. 
The length of one cell is set as $7.5$ m.

In the MLSOV model, the time evolution of $v_{i}^{t}$ is determined not only by $i$-th and $i+1$-th vehicle but also $j+1$-th vehicle 
which is the closest vehicle to $i$-th vehicle on the neighboring lane, with the distance $\Delta x_{2,i}^{t}$ cells ahead 
as shown in figure \ref{fig:mlsov_model}.
$v_{i}^{t}$ is given as
\begin{align}
v_{i}^{t+1}-v_{i}^{t} = a \left\{ V(\Delta x_{1i}^{t},\Delta x_{2i}^{t})-v_{i}^{t}\right\},
\label{eq:v_i_t+1}
\end{align}
where $V(\Delta x_{1,i}^{t},\Delta x_{2,i}^{t})$ is the two-lanes Optimal Velocity (OV) function \cite{M.Bando}.
$v_{i}^{t}$ tends to approach $V(\Delta x_{1,i}^{t},\Delta x_{2,i}^{t})$ with the increase of the sensitivity parameter $a$ ($0 \le a \le 1$).
We define for convenience $\Delta x_{1,i}^{t} = \infty$ in the case that $i+1$-th car does not exist 
and $\Delta x_{2,i}^{t} = \infty$ in the case that $j+1$-th car does not exist.
 
Now, we set the concrete form of $V$ as
\begin{align}
V(\Delta x_{1i}^{t},\Delta x_{2i}^{t}) = \begin{cases}
              0,   & \text{$\Delta x_{1,i}^{t} = 0$} \\
              r,   & \text{$\Delta x_{1,i}^{t} \ge 1$ and $\Delta x_{2,i}^{t} = 0$} \\
              q,   & \text{$\Delta x_{1,i}^{t} \ge 1$ and $\Delta x_{2,i}^{t} = 1$} \\
              p,   & \text{$\Delta x_{1,i}^{t} \ge 1$ and $\Delta x_{2,i}^{t} \ge 2$},\\
            \end{cases}
\label{eq:V12}
\end{align} 
which is simple enough to achieve the alternative configuration as shown below. 
Here we consider special two cases of the time evolution of $v_{i}^{t}$ with the initial condition of $v_{i}^{0} = p$. 
In the case $a=0$, MLSOV model corresponds to the single-lane Asymmetric Simple Exclusion Process (ASEP) \cite{{B. Derrida}}, 
since each vehicle moves irrespective of neighboring ones and $v_{i}^{t}$ is given as
\begin{align}
v_{i}^{t} = \begin{cases}
              0,   & \text{$\Delta x_{1,i}^{t} = 0$} \\
              p,   & \text{$\Delta x_{1,i}^{t} \ge 1$.}
            \end{cases}
\label{eq:va0}
\end{align}
In the case $a=1$, MLSOV model corresponds to a two-lane version of the Zero Range Process (ZRP) \cite{F. Spitzer}, 
and $v_{i}^{t}$ is given as 
\begin{align}
v_{i}^{t} = \begin{cases}
              0,   & \text{$\Delta x_{1,i}^{t} = 0$} \\
              r,   & \text{$\Delta x_{1,i}^{t} \ge 1$ and $\Delta x_{2,i}^{t} = 0$} \\
              q,   & \text{$\Delta x_{1,i}^{t} \ge 1$ and $\Delta x_{2,i}^{t} = 1$} \\
              p,   & \text{$\Delta x_{1,i}^{t} \ge 1$ and $\Delta x_{2,i}^{t} \ge 2$.} \\
            \end{cases}
\label{eq:va1}
\end{align}
Note that in the case $a=0$, the model is exactly solvable and we can theoretically calculate the stationary state and hence the fundamental diagram. 
In the case $a=1$, if we neglect the effect of the other lane, then we can also solve the model exactly. 
\section{SIMULATIONS}
\begin{figure}[htbp]
\includegraphics[width=1.0\linewidth]{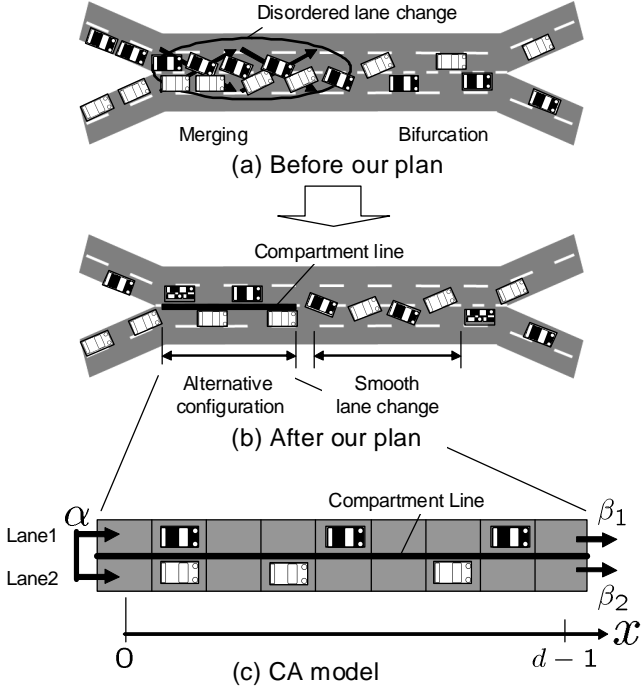}
\caption{
(a) An example of traffic flow on a weaving section before using our plan of drawing a compartment line at the merging area.
Disordered lane changes cause traffic congestion at the merging area. 
(b) An example of traffic flow after introducing our plan. 
The line prohibits vehicles from changing lanes, and is expected to make smooth lane changes by achieving alternative configurations. 
(c)A CA model of a two-lane road with the compartment line. 
Each vehicle enters in the cell at $x=0$ on both lane $1$ and lane $2$ simultaneously with the probability $\alpha$, 
and goes out of the cell at $x=d-1$ with the probability $\beta_{i}$ ($i=1,2$). 
}
\label{fig:no_merging_model}
\end{figure}

Our idea for making the traffic flow smooth is to draw the compartment line 
between the lanes as shown in figure \ref{fig:no_merging_model}. 
The line prohibits vehicles from changing lanes, 
and this will be used for communicating vehicles between lanes and expected to induce the zipper effect at the end of the line. 
Then, the disordered lane change at the merging area will be smooth and becomes safe. 
   
The part of two-lane road indicated in figure \ref{fig:no_merging_model} (b) is partitioned into identical cells 
as shown in figure \ref{fig:no_merging_model} (c). 
The cell size is fixed and one cell can accommodate at most one vehicle. 
The boundaries are open, and each vehicle is updated in parallel, and it enters in the leftmost cell on lane $1$ or lane $2$, 
and moves straight ahead without changing lanes, and goes out of the rightmost cell. 
Although the vehicles cannot change lanes, 
vehicles on one lane can interact with the ones on the opposite lane by looking at the behaviour of these vehicles.

All the parameters are defined in figure \ref{fig:no_merging_model} (c). 
The length of this section is $d$, 
and the leftmost cells are at $x=0$ and the rightmost cells are at $x=d -1$. 

In the simulations we choose the worst injection condition for merging, 
i.e., a vehicle enters in $x=0$ \textit{simultaneously} on lane $1$ and lane $2$ with the probability $\alpha$ 
as long as \textit{both} leftmost cells are empty.
This injection is considered to be the most severe configuration for attaining the alternative configuration at the end of the road where merging occurs. 
We will demonstrate in the following that the alternative configuration is surely achieved in all cases of the injection. 
The probability of going out of the rightmost cell of lane $i$ is set as $\beta_{i}$ ($i=1,2$).

\begin{figure}[htbp]
\includegraphics[width=1.0\linewidth]{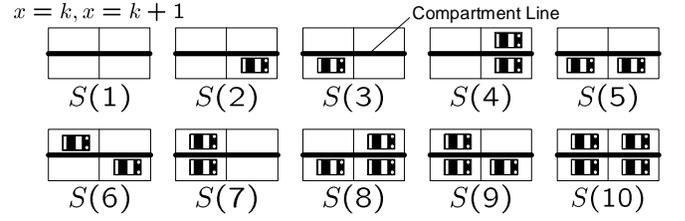}
\caption{ 
$10$ kinds of the state labelled by $S(n)$ in the four cells at $x=\left\{ k,k+1\right\} $. 
The symmetry between lane $1$ and lane $2$ is taken into account to reduce the number of states.
}
\label{fig:state_4cluster}
\end{figure}

\begin{figure}[htbp]
\includegraphics[width=1.0\linewidth]{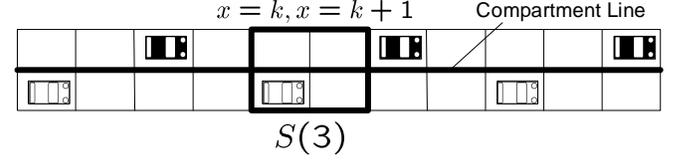}
\caption{ 
The perfect alternative configuration on the two-lane road. 
When at least one vehicle exists at $x=k$, only $S(3)$ represents this perfect alternative configuration. 
}
\label{fig:perfect_alternative_configuration}
\end{figure}

Next, we define two important quantities in the simulations. 
Firstly, we quantify the degree of the alternative configurations of vehicles named \textit{Geminity} (\textit{Ge}).
$Ge$ is a function of $x$, and $Ge(k)$ denotes the degree of the alternative pattern of vehicles at $x=k$. 
In the simulations, $Ge(k)$ is calculated by counting the state of the four cells at $x=\left\{ k, k+1 \right\} $.
There are $10$ kinds of state labelled as $S(n)$ ($n=1,2,...10$) shown as figure \ref{fig:state_4cluster}. 
The symmetry between lane $1$ and lane $2$ is taken into account for eliminating the similar state. 
We defined $c(n)_{k}$ $(n=1,2,...10)$ as the total number of each $S(n)$ at $x=\left\{ k,k+1\right\}$ 
appeared through $M$ times of simulations under the same conditions.
The period for measurement of each simulation is set as $T_{1} \le t \le T_{2}-1$.
When there is at least one vehicle at $x=k$, 
the state of the four cells at $x=\left\{ k,k+1\right\} $ can be $n=3,5,6,7,8,9,10$ in $S(n)$.
Only $S(3)$ among them represents the perfect alternative state which is defined by figure \ref{fig:perfect_alternative_configuration}.
Thus, $Ge(k)$ is given by $c(n)_{k}$ as
\begin{align}
Ge(k) &= c(3)_{k} /\bigl( c(3)_{k} + c(5)_{k} + c(6)_{k} \nonumber \\
      &+ c(7)_{k} + c(8)_{k} + c(9)_{k} + c(10)_{k}\bigr).
\label{eq:Ge_x}
\end{align}
$Ge$ ranges from $0$ to $1$.
The large value of $Ge(k)$ denotes that the alternative configuration of vehicles at $x=k$ is highly achieved. 

Secondly, we define the mean intension, i.e., mean velocity on the two cells at $x=k$ ($0 \le k \le d-1$) denoted by $\bar{v}(k)$, 
which is given by 
\begin{align}
\bar{v}(k) =\frac{\sum_{j=0}^{M-1} \sum_{i=0}^{N_{j}-1} \sum_{t=T_{1}}^{T_{2}-1} v_{i}^{t} \delta_{x_{i}^{t},k}}
{\sum_{j=0}^{M-1} \sum_{i=0}^{N_{j}-1} \sum_{t=T_{1}}^{T_{2}-1} \delta_{x_{i}^{t},k}},
\label{eq:meanv_k_sim}
\end{align}
where $\delta_{x_{i}^{t},k}$ is defined as $1$ if $x_{i}^{t}=k$, and $0$ if $x_{i}^{t}\neq k$. 
This is calculated through $M$ times of simulations under the same conditions, 
and $N_{j}$ ($0 \le j \le M-1$) is the total number of vehicles entering in the leftmost cells on both lane $1$ and lane $2$ in $j$-th simulation. 
The period for measurement of each simulation is the same as that of $Ge$. 

We obtained $Ge(x)$ versus $x$ and  $\bar{v}$ versus $x$ by using numerical simulations. 
The conditions of the simulations are follows. 
Three sets of parameters in the OV function are chosen as $p=1$ and $(q,r)=\left\{(0.99,0.99),(0.8,0.8),(0.5,0.5)\right\}$. 
Moreover, five kinds of the value of the sensitivity parameter are chosen as $a = \left\{0, 0.001, 0.01, 0.1, 1 \right\}$.
The length of the road is set as $d=100$. 
The probability of simultaneous injection on both lane $1$ and lane $2$ is given as $\alpha=0.05$.
Vehicles enter in the leftmost cells with the initial intension $p$ which appears in (\ref{eq:V12}).
$\beta_{i}$ ($i=1,2$) is given as each vehicle's intension $v$ on the cell at $x=d-1$. 
We see that traffic flow is always in a free flow under this $\alpha$ and $\beta_{i}$. 
The number of iterations of the simulations is $M=10$, and the period for measurement is between $T_{1}=100000$ and $T_{2}=200000$. 

The results of the simulations are shown in figure \ref{fig:ge_meanv_sim_4cl}. 
From figure \ref{fig:ge_meanv_sim_4cl} we see that 
$Ge(x)$ increases monotonically from $0$ to $1$ as $x$ became larger in the cases of $a = \left\{0.001, 0.01, 0.1, 1 \right\}$, 
and stays constant at $0$ regardless of $x$ in the cases of $a = 0$.
The sharpness of the increase of $Ge(x)$ becomes larger as $a$ becomes larger, and as $q$($=r$) becomes smaller 
from figure \ref{fig:ge_meanv_sim_4cl} (a)-(c). 
$\bar{v}(x)$ has one minimum value in the case of $a = \left\{0.001, 0.01, 0.1, 1 \right\}$, 
and stays constant at $\bar{v}(x)=1$ in the case of $a=0$ from figure \ref{fig:ge_meanv_sim_4cl} (d)-(f). 
The position $x$ which gives the minimum value of $\bar{v}$ becomes smaller as $a$ becomes larger, and as $q$($=r$) becomes smaller. 
The minimum value of $\bar{v}(x)$ becomes smaller as $a$ becomes larger, and as $q$($=r$) becomes smaller.

Figure \ref{fig:ge_meanv_sim_4cl} (a)-(c) clearly show the achievement of the alternative configuration toward the spatial axis. 
These figures suggest us how we realize the smooth merging at an intersection or a junction. 
The sufficient length of compartment line for communicating is significant for the achievement of the alternative configurations.
For instance, if we set the target $Ge$ as $Ge_{tar} = 0.9$ in the case of $a=0.1$, $p=1$, and $q=0.5$, 
which are typical values of real traffic \cite{R. Nishi Master Thesis},  
then we achieve this value of $Ge$ by drawing the line of $22$ cell length which is equal to $165$ m. 
Note that this alternative configuration is formed only by the local interaction within the range of $\Delta x_{2} \le 1$ of each vehicle. 
The relationships between $a$ and the sharpness of the increase of $Ge(x)$ suggest 
that the length of line necessary for realizing the complete alternative configuration 
becomes shorter if vehicles respond to the ones on the opposite lane more quickly. 
The relationships between $q$ and the sharpness of the increase of $Ge(x)$ suggest 
that this length of line becomes shorter if vehicles decelerate more strongly in the case of $\Delta x_{2} \le 1$. 

The spatial change of $\bar{v}(x)$ is caused by the interactions between the vehicles neighboring each other in two steps.
Firstly, $i$-th vehicle enters into $x=0$ with $\Delta x_{2,i}=0$ 
and slows down due to the OV function $V < 1$ to apart from the ones on the neighboring lane.
It keeps slowing down until $\Delta x_{2,i} \ge 2$. 
Then, $\Delta x_{2,i}=2$ is attained, and it starts to accelerate with $V=1$. 
These interactions generate the minimum value of $\bar{v}(x)$ at the point where $\Delta x_{2}$ of the most vehicle becomes $2$. 
Vehicles slow down more quickly as the response parameter $a$ becomes larger, and $q$($=r$) becomes smaller.
Thus, $x$ which give the minimum value of $\bar{v}(x)$ becomes smaller as $a$ becomes larger, and as $q$ becomes smaller. 
\section{ANALYSIS OF ALTERNATIVE CONFIGURATIONS WITH CLUSTER APPROXIMATION}
In this section, we theoretically study the achievement of alternative configurations by using cluster approximation. 
In the approximation, the two-lane road as shown in figure \ref{fig:no_merging_model} is divided 
into the four cells at $x=\left\{k,k+1\right\}$ $\left(k=0,1,\ldots,d-2\right)$ which is denoted by $C_{k}$. 
Then, the stationary state on $C_{k}$ is calculated in the order of $k=0,1,\ldots,d-2$. 
The state vector of $C_{k}$ at time $t$ is defined as ${\bf\Pi}_{k}^{t}=\left\{ \Pi (1)_{k}^{t},\Pi (2)_{k}^{t},\ldots , \Pi (10)_{k}^{t} \right\}$, 
where $\Pi (n)_{k}^{t}$ ($1 \le n \le 10$) is the probability of $C_{k}$ having the state $S_{n}$ at time $t$ as shown in figure \ref{fig:state_4cluster}. 
The state transition of $C_{k}$ from ${\bf\Pi}_{k}^{t}$ to ${\bf\Pi}_{k}^{t+1}$ is defined 
by using the state transition matrix ${\bf P}_{k}$ whose size is $10 \times  10$. 
This state transition is given as 
\begin{align}
{\bf\Pi}_{k}^{t+1} = {\bf P}_{k} {\bf\Pi}_{k}^{t}.
\label{eq:Pi_k_t_P}
\end{align}
The stationary state ${\bf\Pi}_{k}^{\infty}$ is given as the solution of 
\begin{align}
{\bf\Pi}_{k}^{\infty} = {\bf P}_{k} {\bf\Pi}_{k}^{\infty}
\label{eq:Pi_k_infty}
\end{align}
with the normalized condition $\sum _{n=1}^{10} \Pi(n)_{k}^{\infty} = 1$. 
We construct ${\bf P}_{k}$ to solve (\ref{eq:Pi_k_infty}) in the order of $k=0,1,2,\ldots ,d-2$. 
${\bf P}_{k}$ is given in the following by the boundary conditions (B.C.), and by the dynamics of vehicles. 

\begin{figure}[htbp]
\includegraphics[width=1.0\linewidth]{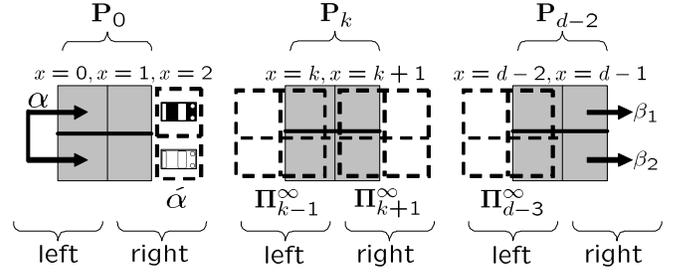}
\caption{
The B.C. of each ${\bf P}_{k}$ ($k=0,1,2,\ldots ,d-2$).
The right B.C. of ${\bf P}_{0}$ is approximated as a pair of vehicles existing on both cells at $x=2$ with probability $\acute{\alpha}$. 
$\acute{\alpha}$ is approximated as $\acute{\alpha}=\alpha/(1+\alpha)$ which is the expected density of the loss system. 
The right B.C. of ${\bf P}_{k}$ ($1 \le k \le d-3$) is the stationary state of $C_{k+1}$ which is approximated by ${\bf\Pi}_{k-1}^{\infty}$. 
}
\label{fig:bc_of_matrix}
\end{figure}

The B.C. of each ${\bf P}_{k}$ ($k=0,1,2,\ldots ,d-2$) are given as shown in figure \ref{fig:bc_of_matrix}. 
The left B.C. of ${\bf P}_{0}$ is given strictly as vehicles entering in the leftmost cells simultaneously with probability $\alpha$ 
provided that both leftmost cells are empty.  
The right B.C. of ${\bf P}_{0}$ needs approximations because the stationary state of $C_{k}$ ($1 \le k \le d-2$) is not yet calculated. 
The right B.C. of ${\bf P}_{0}$ is approximated as a pair of vehicles existing on both cells at $x=2$ with probability $\acute{\alpha}$. 
$\acute{\alpha}$ is approximated as $\acute{\alpha}=\alpha/(1+\alpha)$ which is the expected density of the loss system. 
The left B.C. of ${\bf P}_{k}$ ($1 \le k \le d-3$) is the stationary state of $C_{k-1}$ which is given strictly by ${\bf\Pi}_{k-1}^{\infty}$. 
The right B.C. of ${\bf P}_{k}$ ($1 \le k \le d-3$) is the stationary state of $C_{k+1}$ which is not yet calculated. 
This stationary state is approximated by ${\bf\Pi}_{k-1}^{\infty}$, 
i.e., the stationary state at $x=\left\{ k+1,k+2\right\}$ is approximated as that at $x=\left\{ k-1,k\right\}$. 
This approximation represents well the spatial change of the configurations if the flow is in free state. 
The left B.C. of ${\bf P}_{d-2}$ is the stationary state of $C_{d-3}$ which is given strictly by ${\bf\Pi}_{d-3}^{\infty}$. 
The right B.C. of ${\bf P}_{d-2}$ is given strictly as vehicles go out of the rightmost cell on lane $i$ ($i=1,2$) with probability $\beta_{i}$.   

\begin{figure*}[htbp]
\begin{tabular}{cc}
\begin{minipage}[t]{.5\hsize}
\begin{center}
\includegraphics[width=\hsize,clip]{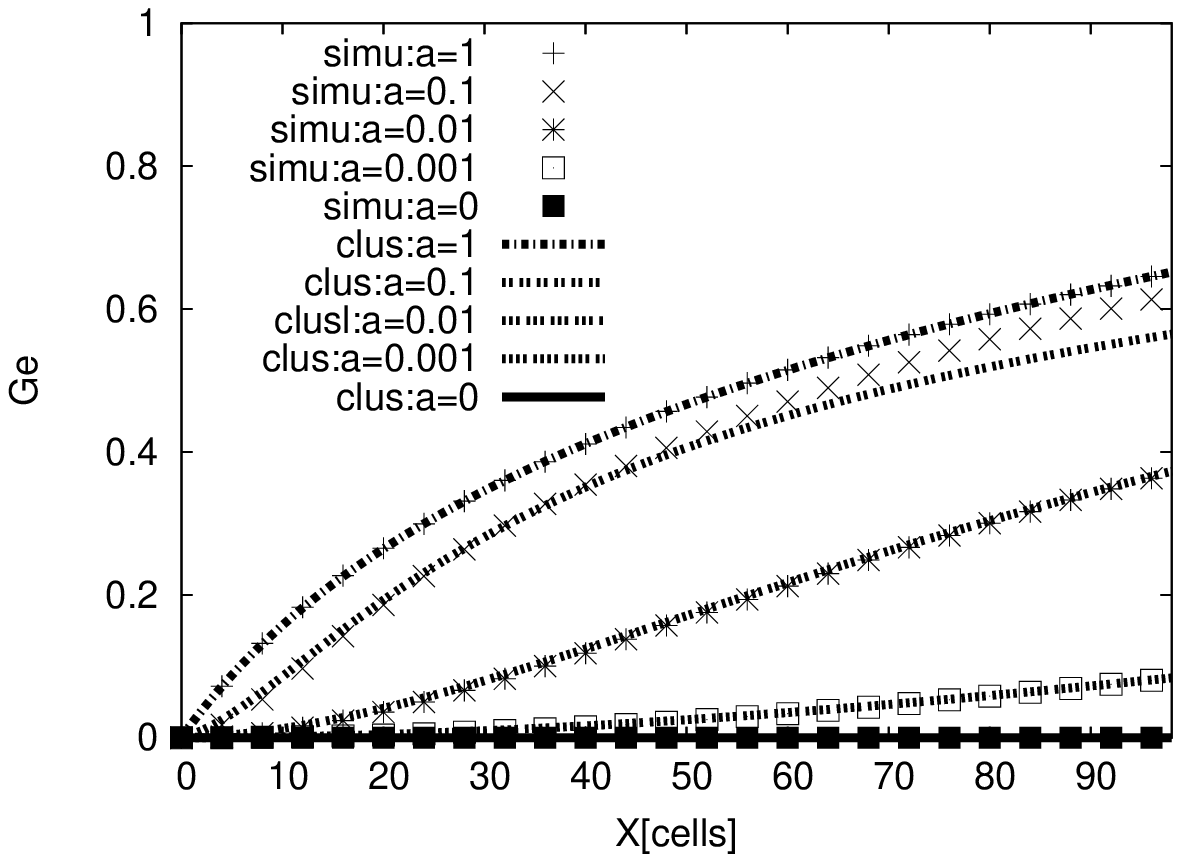}
(a) \textit{Ge} versus $x$, $q=r=0.99$
\label{fig:ge_4cl_sim_099}
\end{center}
\end{minipage}
\begin{minipage}[t]{.5\hsize}
\begin{center}
\includegraphics[width=\hsize,clip]{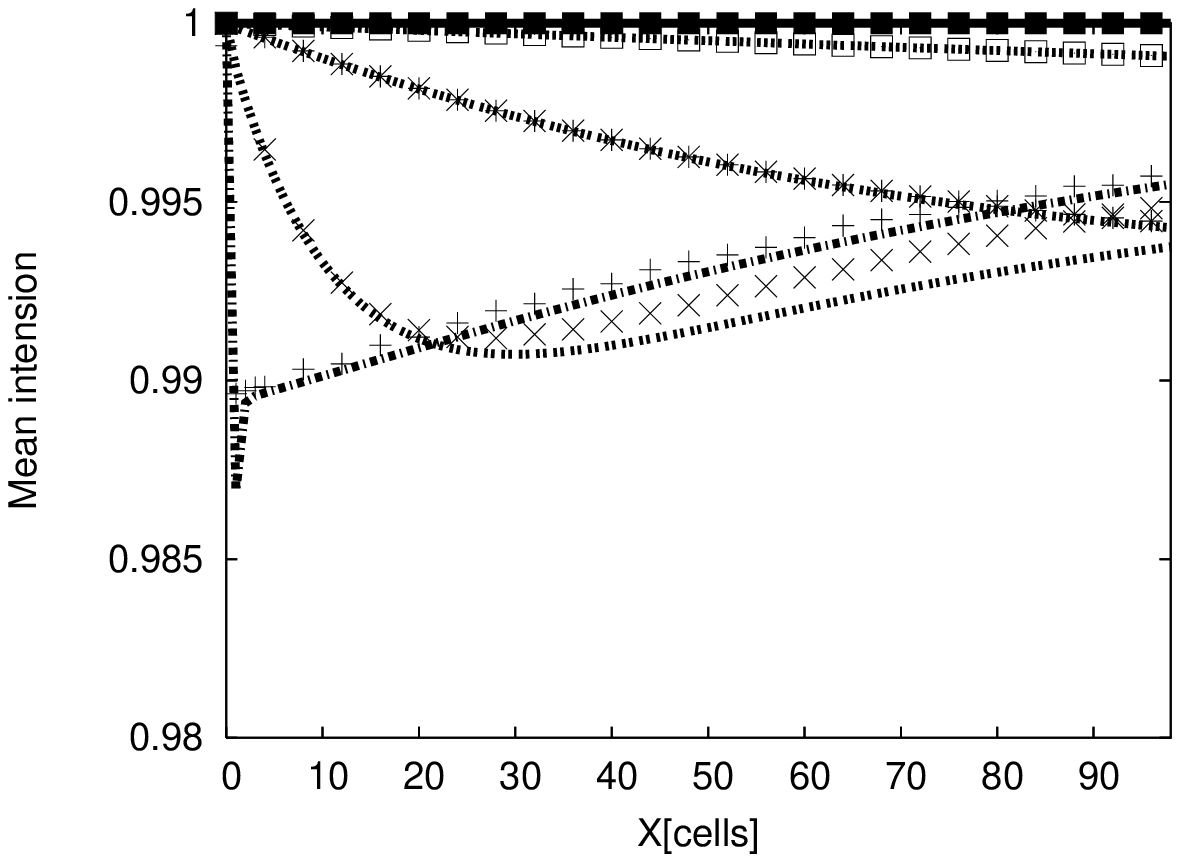}
(d) mean $v$ versus $x$, $q=r=0.99$
\label{fig:v_4cl_sim_099}
\end{center}
\end{minipage}
\end{tabular}
\begin{tabular}{cc}
\begin{minipage}[t]{.5\hsize}
\begin{center}
\includegraphics[width=\hsize,clip]{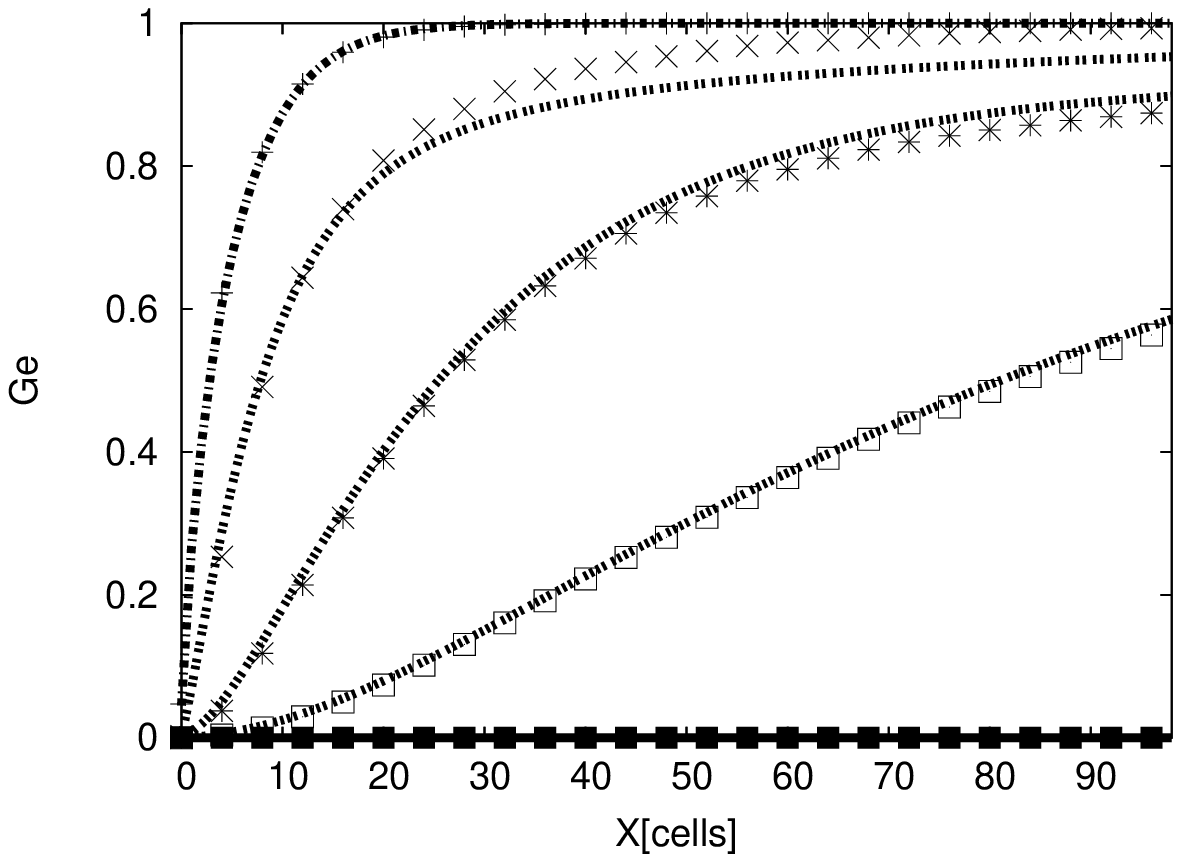}
(b) \textit{Ge} versus $x$, $q=r=0.8$
\label{fig:ge_4cl_sim_08}
\end{center}
\end{minipage}
\begin{minipage}[t]{.5\hsize}
\begin{center}
\includegraphics[width=\hsize,clip]{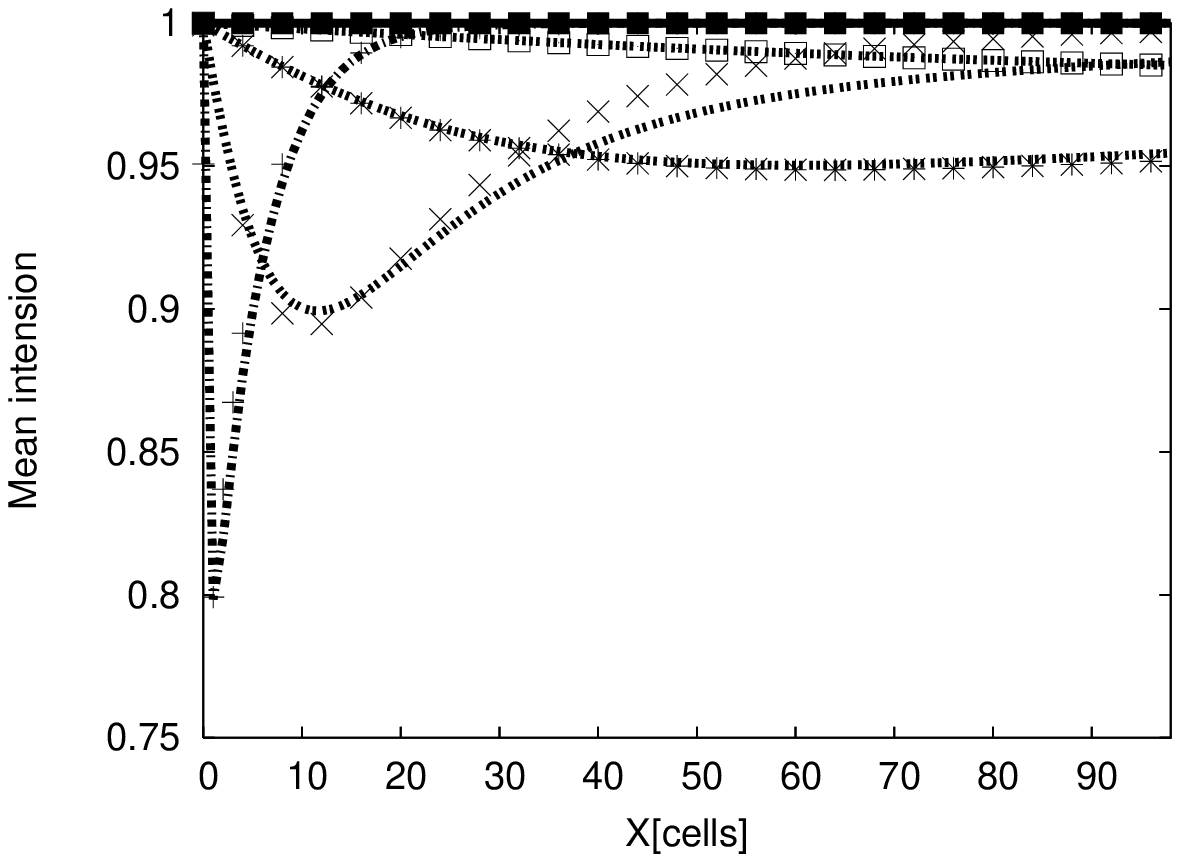}
(e) mean $v$ versus $x$, $q=r=0.8$
\label{fig:v_4cl_sim_08}
\end{center}
\end{minipage}
\end{tabular}
\begin{tabular}{cc}
\begin{minipage}[t]{.5\hsize}
\begin{center}
\includegraphics[width=\hsize,clip]{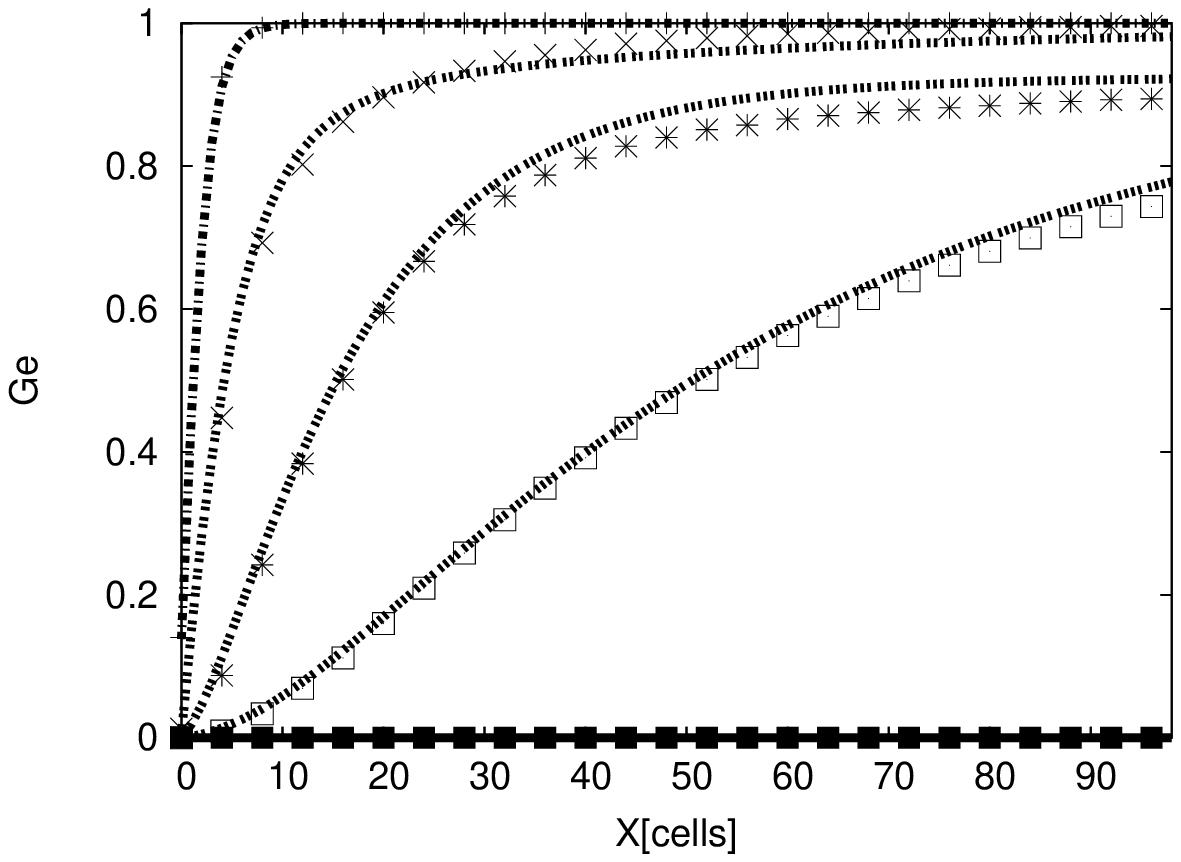}
(c) \textit{Ge} versus $x$, $q=r=0.5$
\label{fig:ge_4cl_sim_05}
\end{center}
\end{minipage}
\begin{minipage}[t]{.5\hsize}
\begin{center}
\includegraphics[width=\hsize,clip]{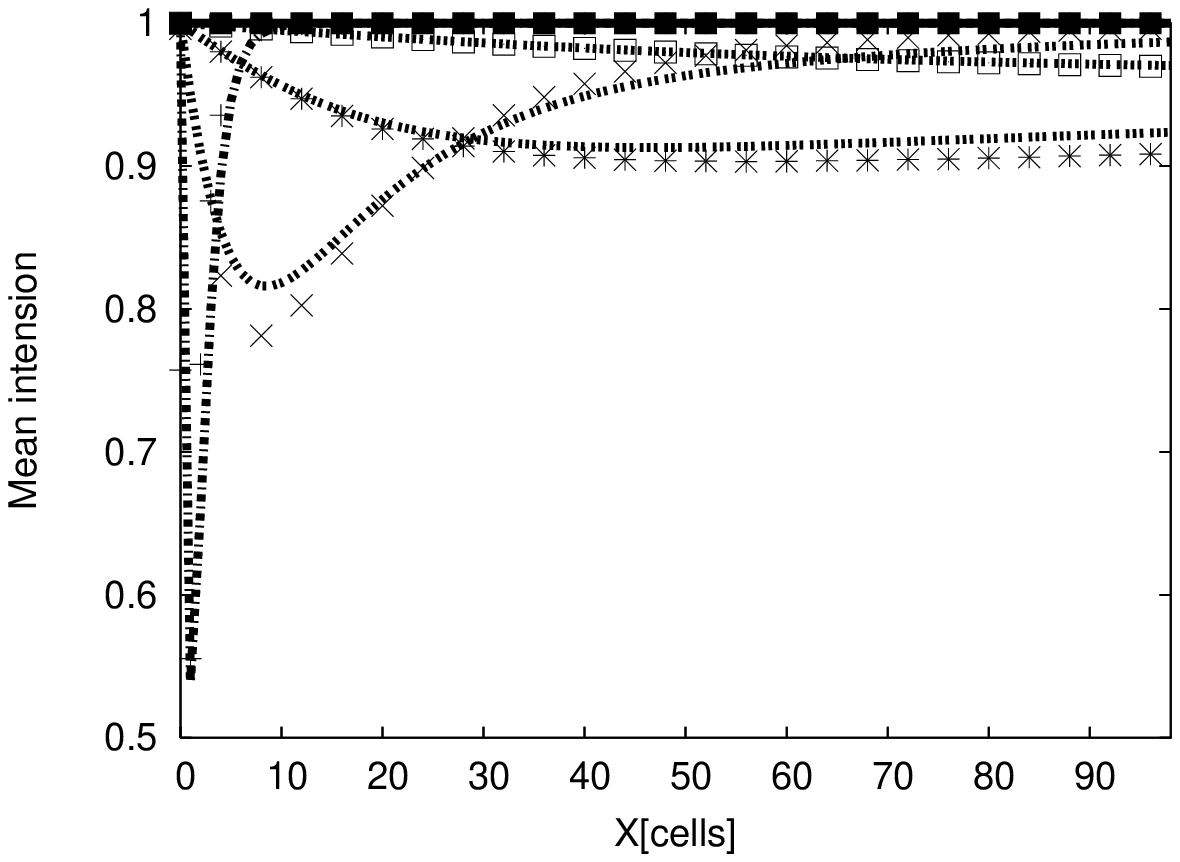}
(f) mean $v$ versus $x$, $q=r=0.5$
\label{fig:v_4cl_sim_05}
\end{center}
\end{minipage}
\end{tabular}
\caption{
$Ge(x)$ versus $x$ and $\bar{v}$($x$) versus $x$ obtained by the simulations, 
together with $Ge_{x}^{\infty}$ versus $x$ and $\tilde{v}_{x}$ versus $x$ calculated by the four cluster approximation. 
$Ge(x)$ and $Ge_{x}^{\infty}$ increase monotonically as $x$ increases, 
and the sharpness of the increase of $Ge(x)$ and $Ge_{x}^{\infty}$ becomes larger as $a$ becomes larger, 
and as $q$(=$r$) becomes smaller.
$\bar{v(x)}$ and $\tilde{v}_{x}$ have one minimum value.
}
\label{fig:ge_meanv_sim_4cl}
\end{figure*}

The dynamics of vehicles in this cluster approximation is given as follows.                  
We update the intension of vehicles spatially to represent the spatial change of the mean intension $\bar{v}$ 
observed in the simulation results (figure \ref{fig:ge_meanv_sim_4cl} (d)-(f)).
To realize the spatial change of the intension, we define $u_{i,k}$ $\left(0 \le k \le d-2 \right)$ as the intension of $i$-th vehicle on $C_{k}$. 
To give the specific form of $u_{i,k}$, we use $\tilde{v}_{k}$ defined as the intension common to the vehicles on $C_{k}$, 
and $\bar{V}_{k}$ ($0 \le k \le d-2$) defined as the stationary mean OV function of $C_{k}$.
$\bar{V}_{k}$ is calculated by using ${\bf\Pi}_{k}^{\infty}$ as
\begin{align}
\bar{V}_{k} = &(p \Pi (3)_{k}^{\infty} + q \Pi (6)_{k}^{\infty} + 2r \Pi (7)_{k}^{\infty} + r \Pi (9)_{k}^{\infty})/ \nonumber \\
              &(\Pi (3)_{k}^{\infty} + \Pi (5)_{k}^{\infty} + \Pi (6)_{k}^{\infty} + 2 \Pi (7)_{k}^{\infty} + \nonumber \\
              &\Pi (8)_{k}^{\infty} + 2 \Pi (9)_{k}^{\infty} + 2 \Pi (10)_{k}^{\infty}),
\label{eq:bar_V}
\end{align}
and $\tilde{v}_{k}$ is updated spatially as
\begin{align}
\tilde{v}_{k+1} = (1-a) \tilde{v}_{k} + a \bar{V}_{k}.
\label{eq:v_tilde_update}
\end{align}
$u_{i,k}$ is given by using $\tilde{v}_{k}$, and by the configuration on $C_{k}$ as
\begin{align}
u_{i, k}   &= (1-a) \tilde{v}_{k} + a V(\Delta x_{1i},\Delta x_{2i}) \nonumber \\
          &=  \begin{cases}
              (1-a) \tilde{v}_{k},         & \text{$\Delta x_{1i} = 0$} \\
              (1-a) \tilde{v}_{k} + a r,   & \text{$\Delta x_{1i} \ge 1$ and $\Delta x_{2i} = 0$} \\
              (1-a) \tilde{v}_{k} + a q,   & \text{$\Delta x_{1i} \ge 1$ and $\Delta x_{2i} = 1$} \\
              (1-a) \tilde{v}_{k} + a p,   & \text{$\Delta x_{1i} \ge 1$ and $\Delta x_{2i} \ge 2$.}\\
              \end{cases}
\label{eq:u_ik}
\end{align}

Now the concrete form of $P_{k}$ is given as follows. 
We define $P_{k}\left(i,j\right)$ $\left(1 \le i,j \le 10\right)$ as the element of ${\bf P}_{k}$ at $i$-th row, and $j$-th column. 
$P_{k}\left(i,j\right)$ is the probability of the transition of the state from $S(j)$ to $S(i)$.  
For instance, $P_{k}\left(1,1\right) = 1-\alpha$, and $P_{k}\left(7,1\right) = \alpha$. 
The other elements of $P_{0}$, and $P_{k}$ ($1 \le k \le d-3$), 
and $P_{d-2}$ obeying the B.C. of the cells at $x=d-1$, such that $\beta_{j}=u_{i, d-1}$ ($j=1,2$) for $i$-th vehicle,
are straightforward and shown in \cite{R. Nishi Master Thesis}. 

The degree of the alternative configurations at $x=k$ ($0 \le k \le d-2$) in the stationary state is defined 
as $Ge_{k}^{\infty}$ which is given by using ${\bf\Pi}_{k}^{\infty}$ as
\begin{align}
Ge_{k}^{\infty} &= \Pi (3)_{k}^{\infty}/ \bigl( \Pi (3)_{k}^{\infty} + \Pi (5)_{k}^{\infty} + \Pi (6)_{k}^{\infty} \nonumber \\
                &+ \Pi (7)_{k}^{\infty} + \Pi (8)_{k}^{\infty} + \Pi (9)_{k}^{\infty} + \Pi (10)_{k}^{\infty} \bigr) .
\label{eq:ge_4clu}
\end{align}

We measure $Ge_{x}^{\infty}$ versus $x$, and $\tilde{v}_{x}$ versus $x$, and compare them with the simulation results. 
The conditions of the cluster approximation are given similarly to the simulations. 
$d$, $p$, $q$, $r$, $a$, and $\alpha$ are the same to those of the simulations. 
$\beta_{j}$ ($j=1,2$) is given as the $u_{i,d-1}$ for each $i$-th vehicle.

The results of the four cluster approximation are shown in figure \ref{fig:ge_meanv_sim_4cl} 
together with the simulation results to compare the theoretical results with the simulations.

Both $Ge_{x}^{\infty}$ and  $\tilde{v}_{x}$ coincide with $Ge$($x$), and $\bar{v}$($x$) respectively.
The difference $\left|Ge_{x}^{\infty}-Ge(x)\right|$ is smaller in the case of $a=0$ and $a=1$ than that of $a=0.1$. 

The coincidence between $Ge_{x}^{\infty}$ and $Ge$($x$) with various $a$ and $q$ suggests 
that our four cluster approximation is a good theoretical approximation for obtaining the alternative formation. 
We can obtain the length of the communication line necessary for the target of $Ge$ by the theoretical calculation as well as by the simulations. 
Various $Ge_{x}^{\infty}$ are generated by the spatial update of the intension as shown in (\ref{eq:bar_V})-(\ref{eq:u_ik}).
$\left|Ge_{x}^{\infty}-Ge\left(x\right)\right|$ in the case of $a=0$ and $a=1$ are smaller than that of $a=0.1$ 
because the calculation of $u_{i, k}$ does not contain the approximation of the spatial update of $\tilde{v}_{x}$ in the case of $a=0$ and $a=1$: 
$u_{i,k}$ in the case of $a=0$ is given as
\begin{align}
u_{i, k} = \tilde{v}_{0} 
\label{eq:u_ik_a_0}
\end{align}
and $u_{i,k}$ in the case of $a=1$ is given as
\begin{align}
u_{i, k}  =  \begin{cases}
              0,     & \text{$\Delta x_{1i} = 0$} \\
              r,     & \text{$\Delta x_{1i} \ge 1$ and $\Delta x_{2i} = 0$} \\
              q,     & \text{$\Delta x_{1i} \ge 1$ and $\Delta x_{2i} = 1$} \\
              p,     & \text{$\Delta x_{1i} \ge 1$ and $\Delta x_{2i} \ge 2$.}\\
              \end{cases}
\label{eq:u_ik_a_1}
\end{align}

The coincidence of $\tilde{v}_{x}$ and $\bar{v}$($x$) suggests 
that this spatial update of the intension is a good approximation for expressing the spatial change of the mean intension. 
\section{CONCLUDING DISCUSSIONS}
In this paper we have proposed a simple method for the achievement of alternative configurations of vehicles by using MLSOV model.
We obtain for the first time the achievement of alternative configurations of vehicles 
by emergent behaviour of drivers.
This alternative configurations is achieved along a compartment line 
which is drawn between two lanes, and prohibits vehicles from changing lanes, and permits them to interact with the ones on the opposite lane. 
This achievement is significant for traffic engineering in realizing the smooth "zipper merging", 
and is supported by the four cluster approximation. 
Moreover, drawing compartment lines for realizing this alternative configuration costs less 
than other methods for easing traffic congestion at an intersection, e.g., construction of a cubic interchange.  
Further improvement of the flow at an intersection or a junction by the alternative configuration 
will be published elsewhere \cite{R. Nishi}. 
\newpage

\end{document}